\documentclass{article}

 \usepackage[final]{neurips_2018}

\usepackage[utf8]{inputenc} 
\usepackage[T1]{fontenc}    
\usepackage[bookmarks=false]{hyperref}       
\usepackage{url}            
\usepackage{booktabs}       
\usepackage{amsfonts}       
\usepackage{nicefrac}       
\usepackage{microtype}      

\usepackage{amsmath}
\usepackage{multirow}
\usepackage{tabularx}
\usepackage{caption}
\usepackage{array}
\usepackage{pbox}
\usepackage{graphicx}
\usepackage{bm}
\usepackage{tikz}

\usepackage[numbers, sort&compress]{natbib}
\usepackage{booktabs, tabularx}

\title{Tracking-Assisted Segmentation of Biological Cells}

\newcommand{\ct}[1]{{\protect\NoHyper\textit{\citeauthor{#1}  \citeyear{#1}}\protect\endNoHyper} [\citenum{#1}]}
\renewcommand\citet{\ct}

\author{%
  Deepak K. Gupta$^*$\\
  University of Amsterdam\\
  \texttt{d.k.gupta@uva.nl} \\
  \And
   Nathan de Bruijn\thanks{Equal contribution.}\\
  University of Amsterdam \\
  \texttt{nathanldebruijn@gmail.com} \\
  \And
  Andreas Panteli\\
  University of Amsterdam \\
  \texttt{andreas.panteli@student.uva.nl} \\
  \And
  Efstratios Gavves\\
  University of Amsterdam \\
  \texttt{e.gavves@uva.nl} \\
}

\begin{document}

\maketitle

\begin{abstract}
U-Net and its variants have been demonstrated to work sufficiently well in biological cell tracking and segmentation. However, these methods still suffer in the presence of complex processes such as collision of cells, mitosis and apoptosis. In this paper, we augment U-Net with Siamese matching-based tracking and propose to track individual nuclei over time. By modelling the behavioural pattern of the cells, we achieve improved segmentation and tracking performances through a re-segmentation procedure. Our preliminary investigations on the Fluo-N2DH-SIM+ and Fluo-N2DH-GOWT1 datasets demonstrate that absolute improvements of up to 3.8 \% and 3.4\% can be obtained in segmentation and tracking accuracy, respectively.
\end{abstract}

\section{Introduction}

\begin{figure}[h]
	\centering
	\includegraphics[scale=0.18]{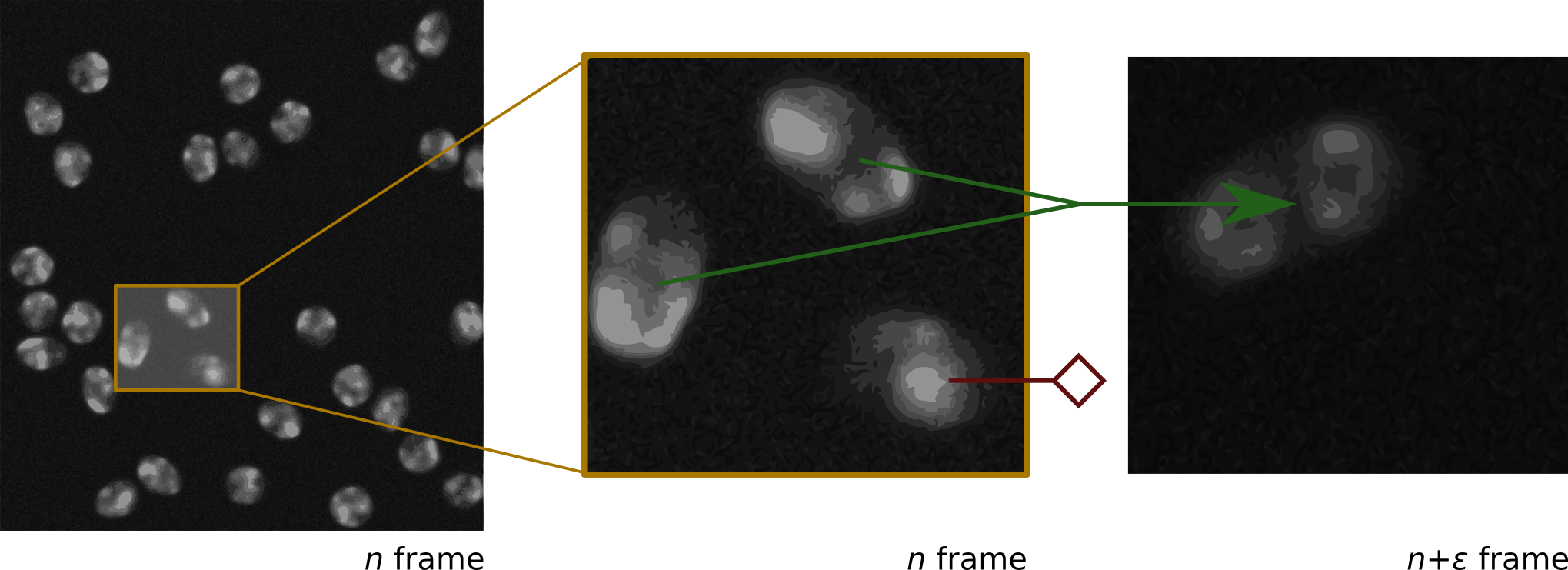}
	\caption{Schematic representation of cell states in two different frames (modified from an image of HL60 cells from the Fluo-N2DH-SIM+ dataset). One cell fades out (cell death) and two collide, in a later frame.}
	\label{fig:sample}
\end{figure}

Small sized biomedical data, such as images of biological cells or lymphocytes, are often difficult to acquire and even harder to visualise due to the restricting scale of particles, the low resolution scans and the viscosity of moving cells \cite{saltz2018spatial, swiderska2019learning}. As illustrated in Figure \ref{fig:sample}, the features of the nuclei in each frame are not always clearly visible and their position is very volatile. Being able to detect individual cells and track their trajectory through time will help automate treatment observation and disease spread detection \cite{coudray2018classification}.

\citet{unet} introduced the U-Net architecture, which has demonstrated state-of-the-art performance on many biomedical image segmentation tasks \cite{unet, falk2019u}. Since then, several cell tracking approaches have utilised the success of U-Net to boost their performances \cite{li2018h}. However, due to the constant change of the position, shape and status of the cells in the data, most such approaches fail to accurately detect cells that combine, split or die (leave the image) \cite{christ2016automatic}.

In this paper, we propose to improve the segmentation of biological cells by augmenting with Siamese matching. The initial U-Net segmentation results are combined with cell detection from a Siamese matching-based tracker \cite{tao2016siamese} to improve the cell recognition in subsequent frames. Next, a dedicated detection mechanism for cell collisions, mitosis (splitting of cells) and apoptosis (cell death) is introduced which attempts to model the movement behaviour of the nuclei. Based on the corrections, we re-segment the cells using random walker algorithm. \cite{grady2006random}. 
\vspace{-0.2em}
\section{Approach}

\begin{figure}
	\centering
	\includegraphics[width=\linewidth]{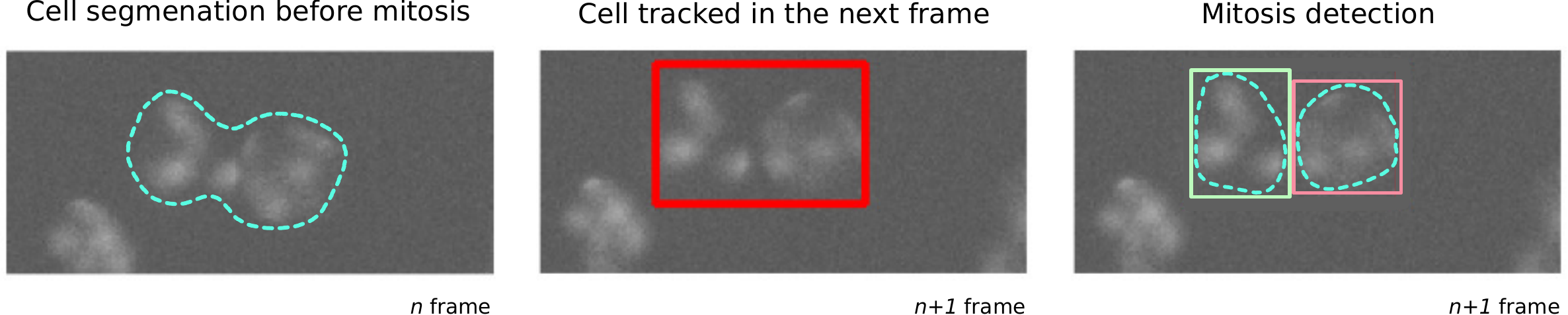}
	\caption{An illustration of Mitosis detection through cell tracking}
	\label{fig:tracking}
	\vspace{-0.5em}
\end{figure}

We propose to use Siamese matching approach to detect the cell collisions as well as mitosis. 
Our model combines the U-Net segmentation with an object tracking mechanism in order to improve upon the initial predictions. The various steps are described below.

\textbf{Initial Segmentation.} 
The initial segmentation is done using the U-Net implementation of \cite{unet}, and the results are used as the baseline segmentation for the chosen sequence. During training, data augmentation consisting of random flips and shifts is employed, and as a final step, a nearest neighbour interpolation algorithm is used convert the U-Net results of $512 \times 512$ pixels to the desired resolution. After this process is finished, cells are detected and defined as being connected regions of positively labelled pixels in the segmentation map.

\textbf{Tracking.} To predict the new location of a cell in a consecutive frame, we use the Siamese tracker of \cite{siamfc} and use a pre-trained model, without any further training on cellular footage. SiamFC is adapted for grayscale images and a search space of $150\times 150$ pixels is used. Tracking is done in the forward as well as the backward directions. During tracking, given that our segmentation and location predictions are independent, we refine the tracking performance and detect the occurrence of mitosis and collision events.

Let $I_t$ denote the $t^{\text{th}}$ frame in a sequence of length $T$, and $\mathcal{S}_t = \{ C_t^1,...,C_t^K\}$ be the set of detected cells in this frame. These are used to initialize the tracker at step $t$. For cell $C^i_t$, we refer to the predicted locations by the tracker in $I_{t+1}$ and $I_{t-1}$ as forward $(F_t^i)$ and backward $(B^i_t)$ predictions, respectively. Collision and mitosis are then detected as follows.

\textbf{Collision detection.} Collision refers to scenarios where two cells share a fraction of their boundary, and this can often be mistaken as a single cell during segmentation. When processing a new frame $I_t$, where $t > 1$, we start off by performing collision detection in which a cell $C_{t}^i$ is considered to be a lump of multiple individual cells if the centroids of two or more cells in $\mathcal{S}_{t-1}$ lie within the tracked region $B_{t}^i$. If this is the case, $C_{t}^i$ is re-segmented according to a procedure based on Random Walker algorithm. More details on this are described later in Re-segmentation section. This collision detection procedure continues until each cell in $I_t$ matches at most one cell in $I_{t-1}$ or until the re-segmentation procedure fails and does not yield improvement anymore.

\textbf{Mitosis detection.} We then continue by matching cells in $\mathcal{S}_{t-1}$ to the detected cells in $I_{t}$. This matching happens in a manner similar to the collision detection, namely a cell $C_{t-1}^i$ is matched to a cell $C_{t}^i$ if the centroid of $C_{t}^i$ is inside the region $F_{t-1}^i$. Different from collision detection, however, $C_{t-1}^i$ is also matched to $C_{t}^i$ if the centroid of the region $F_{t-1}^i$ lies within the boundaries of the cell $C_{t}^i$. This matching procedure yields a set of matches for each cell $C_{t-1}^i$, which we denote as $M_{t-1}^i$ and its size as $|M_{t-1}^i|$. The state of =cell $C_{t-1}^i$ is then determined according to:

\begin{align}
    \label{continuation}
    C_{t-1}^i-\text{state} &= 
    \begin{cases}
        \text{Apoptosis}, &  |M_{t-1}^i| = 0\\
        M_{t-1,1}^i,   &  |M_{t-1}^i| = 1\\
        \text{Mitosis}, & \text{otherwise}
    \end{cases}
\end{align}

In case of mitosis, the cell splits, thus the tracking of $C_{t-1}^i$ ends and the cells in $|M_{t-1}^i|$ are initialised with two new trackers which have $C_{t-1}^i$ as their parent. After continuations have been determined for all cells in $\mathcal{S}_{t-1}$, cells in $\mathcal{S}_{t}$ that are not linked to any cell in $I_{t-1}$ are interpreted as newly detected cells which start their life in $I_{t}$ without link to a parent cell. An illustration is shown in Figure \ref{fig:tracking}.

\begin{figure}
	\centering
	\includegraphics[scale=0.4, trim=0 2cm 0 0.0cm, clip=true]{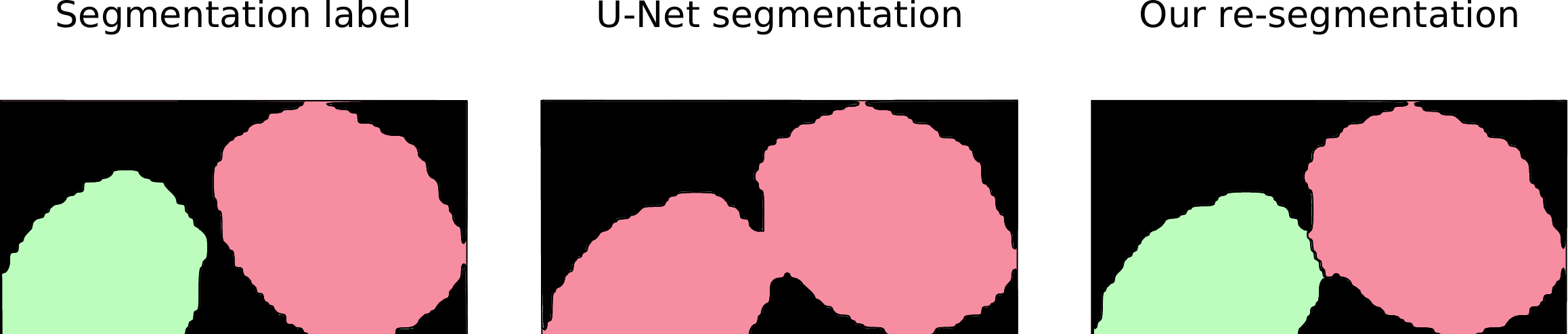}
	\caption{Cell re-segmentation over initial U-Net segmentation}
	\label{fig:segmentation}
\end{figure}

\textbf{Re-segmentation.} In case of a detected collision of two or more cells into a cell $C_{t}^i$, we re-segment  $C_{t}^i$ in such a manner that the new number of segments matches the number of colliding cells. This is achieved using the random walker segmentation algorithm as described in \cite{grady2006random}. To prevent over-segmentation of the cell $C_{t}^i$, which adversely affects segmentation accuracy, we use the relative position of the centroids of the cells in $I_{t-1}$ as the seeds for the segmentation algorithm. An illustration of re-segmented cells is shown in Figure \ref{fig:segmentation}.

\vspace{-0.2em}
\section{Results}
The results of this work are compared against the vanilla U-Net, \cite{unet}, as the baseline approach. The accuracy scores, as described in \cite{CTC}, are used as the decisive metric for model performance. The segmentation and tracking performance of the two methods is listed in Table \ref{tab:seg_results}. As can be seen, the method proposed outperforms the baseline approach on both datasets in both segmentation and tracking with a maximum increase in performance of 3.8\% and  3.4\% on the segmentation and tracking of the Fluo-N2DH-SIM+ 02 collection, respectively.

The increased performance of the method proposed by this work indicates that the mitosis, apoptosis and cell fusion interpretation of the collision detection mechanism indeed help the tracker network improve the segmentation performance. Furthermore, this finding highlights the importance of modelling the movement behaviour of the cells to better capture the pattern nature of cells in consecutive frames. 

\begin{table}
    \centering
    \begin{tabular}{l |c | c || c |c }
     \multicolumn{1}{c}{} & \multicolumn{2}{c||}{Segmentation accuracy} & \multicolumn{2}{c}{Tracking accuracy} \\
     \toprule
     \multicolumn{1}{c|}{\textbf{Dataset}} & \textbf{U-Net} & \textbf{Our method} & \textbf{U-Net + SiamTracker} & \textbf{Our method}\\
     \midrule
     Fluo-N2DH-SIM+ 01   & 0.919 & \textbf{0.924} & 0.986 & \textbf{0.992} \\
     Fluo-N2DH-SIM+ 02   & 0.800 & \textbf{0.838} & 0.922 & \textbf{0.956} \\
     \toprule
     Fluo-N2DH-GOWT1 01  & 0.739 & \textbf{0.746} & 0.973 & \textbf{0.978}\\
     Fluo-N2DH-GOWT1 02  & 0.827 & \textbf{0.837}  & 0.869 & \textbf{0.875}\\
     \hline
    \end{tabular}
    \caption{Accuracy results for segmentation and tracking of cells. Our method  includes the modules for detecting collisions, mitosis and apoptosis on top of U-Net and Siamese matching.}
    \label{tab:seg_results}
    \vspace{-1.7em}
\end{table}

\vspace{-0.2em}
\section{Conclusions}
Deep learning trackers such as U-Net work well on biological cell tracking problems. However, they still suffer in the presence of events such as cell collisions, mitosis and apoptosis. In this work, we proposed to combine Siamese matching \cite{tao2016siamese, siamfc} with U-Net \cite{unet} to accurately identify these difficult scenarios. Siamese matching-based detection efficiently provides a more precise understanding and modelling of individual cells, thereby providing better grasp of latent information about them. Based on the proposed approach, we obtained absolute improvements of up to 3.8 \% and 3.4\% for segmentation and tracking accuracy, respectively. Our future work includes rigorously testing the proposed methodology on a more diverse set of cell tracking data.

\bibliography{egbib_temp}
\bibliographystyle{unsrtnat}
\end{document}